\documentclass[pre,superscriptaddress,twocolumn]{revtex4}
\usepackage[dvips]{graphicx}
\usepackage{amsmath,amsthm,amssymb,subfigure,rotating,color}
\usepackage{longtable}

\begin{document}
\title{Temporal Evolution of Financial Market Correlations}
\author{Daniel J. Fenn}
\affiliation{Mathematical and Computational Finance Group, Mathematical Institute, University of Oxford, Oxford OX1 3LB, UK}
\affiliation{CABDyN Complexity Centre, University of Oxford, Oxford OX1 1HP, UK}
\affiliation{FX Quantitative Strategy, HSBC Bank, 8 Canada Square, London E14 5HQ, UK}
\author{Mason A. Porter}
\affiliation{Oxford Centre for Industrial and Applied Mathematics, Mathematical Institute, University of Oxford, Oxford OX1 3LB, UK}
\affiliation{CABDyN Complexity Centre, University of Oxford, Oxford OX1 1HP, UK}
\author{Stacy Williams}
\affiliation{FX Quantitative Strategy, HSBC Bank, 8 Canada Square, London E14 5HQ, UK}
\author{Mark McDonald}
\affiliation{FX Quantitative Strategy, HSBC Bank, 8 Canada Square, London E14 5HQ, UK}
\author{Neil F. Johnson}
\affiliation{Department of Physics, University of Miami, Coral Gables, Florida 33146, USA}
\affiliation{CABDyN Complexity Centre, University of Oxford, Oxford OX1 1HP, UK}
\author{Nick S. Jones}
\affiliation{Department of Physics, Clarendon Laboratory, University of Oxford, Oxford OX1 3PU, UK}
\affiliation{Oxford Centre for Integrative Systems Biology, University of Oxford, Oxford OX1 3QU, UK}
\affiliation{CABDyN Complexity Centre, University of Oxford, Oxford OX1 1HP, UK}

%%%%%%%%%%%%%%%%

\begin{abstract}

We investigate financial market correlations using random matrix theory and principal component analysis. We use random matrix theory to demonstrate that correlation matrices of asset price changes contain structure that is incompatible with uncorrelated random price changes. We then identify the principal components of these correlation matrices and demonstrate that a small number of components accounts for a large proportion of the variability of the markets that we consider. We characterize the time-evolving relationships between the different assets by investigating the correlations between the asset price time series and principal components. Using this approach, we uncover notable changes that occurred in financial markets and identify the assets that were significantly affected by these changes. We show in particular that there was an increase in the strength of the relationships between several different markets following the 2007--2008 credit and liquidity crisis.

\end{abstract}

\maketitle

%%%%%%%%%%%%%%%%%%%%%%%%%%%%%%%%%%%%%%%%%%%%%%%%%%%%%%%%%%%%%%%%%%%

\section{Introduction}

The global financial system is composed of a variety of markets, which are spread across multiple geographic locations and in which a broad range of financial products are traded. Despite the diversity of markets and the disparate nature of the products that are traded, price changes of assets often respond to the same economic announcements and market news \cite{ederington,balduzzi,andersen}. The fact that asset prices depend on the same signals implies that there is strong coupling between prices, so price time series can exhibit similar characteristics and be correlated. One of the primary concerns of market practitioners is to estimate the strength of such correlations \cite{Laloux_PRL_1999}.

There are many reasons for wanting to understand correlations in price movements. Perhaps the most familiar motivation is for risk management purposes, because large changes in the value of a portfolio are more likely if the prices of the assets held in the portfolio are correlated \cite{Laloux_PRL_1999}. An understanding of the correlation between financial products is therefore crucial for managing investment risk. It has also been shown that the strength of the correlations between some markets can be explained by macroeconomic factors \cite{yang_macro,li_macro}. An understanding of correlations can therefore illuminate the macroeconomic forces driving markets and help inform asset allocation decisions \cite{li_macro}.

In this paper, we use principal component analysis (PCA) to produce a parsimonious representation of market correlations and characterize the evolving correlation structures within markets. PCA is an established tool in data analysis for generating lower-dimensional representations of multivariate data \cite{pcajolliffe} and has provided useful insights in a diverse range of fields \cite{PCA_chemistry,PCA_genetics,PCA_psychology,PCA_astro}. In finance, PCA has been used to identify common factors in international bond returns \cite{driesson,perignon} and produce market indices \cite{pcaindex}.  It has also been used in subjects such as arbitrage pricing theory \cite{chamberlain,connorAPT} and portfolio theory \cite{multicurr}.

PCA is closely linked to random matrix theory (RMT), which was developed to deal with the statistics of the energy levels of many-body quantum systems \cite{wigner,Mehta_RMT}. The standard financial application of RMT is to compare the eigenvalues and eigenvectors of correlation matrices of asset returns with the corresponding properties of correlation matrices for randomly distributed returns \cite{Laloux_PRL_1999,Plerou_PRL_1999,Plerou_PRE_2002,bomm,zum,kenn}. In prior studies, most of the eigenvalues of market correlation matrices were found to lie within the ranges predicted by RMT (for example, in Ref.~\cite{Laloux_PRL_1999}, 94\% of the eigenvalues lie within the RMT range), which is usually taken as an indication that to a large extent the correlation matrix is random and dominated by noise \cite{Laloux_PRL_1999,Plerou_PRL_1999,Plerou_PRE_2002}. In addition, the smallest eigenvalues of market correlation matrices were found to be most sensitive to noise in the estimation of the correlation coefficients \cite{Laloux_PRL_1999}. Because the eigenvectors corresponding to the smallest eigenvalues are used to determine the least risky portfolios in Markowitz portfolio theory \cite{markowitz}, this result has important implications for risk management \cite{Laloux_PRL_1999,Plerou_PRL_1999,Plerou_PRE_2002,Kim_PRE_2005}.

Most prior investigations of markets using PCA and RMT focused on specific markets. For example, there is a large body of work investigating equity markets \cite{Laloux_PRL_1999,Plerou_PRL_1999,Plerou_PRE_2002,Kim_PRE_2005}, and there have also been investigations of emerging market equities \cite{wilcox_EM,cukur_EM,pan_EM}, the foreign exchange (FX) market \cite{drozdz_FX}, and bond markets \cite{driesson,perignon}. Many of these studies only consider a single correlation matrix, although some also investigate the temporal evolution of correlations \cite{bomm,zum}. The work in this paper differs from these studies by investigating a diverse range of asset classes and by studying the time evolution of the correlations between these assets. By studying the time dynamics of the correlations of many different markets, we uncover periods during which there were major changes in the correlation structure of the global financial system. We find that there was a large increase in the strength of correlations between many different assets following the 2007--2008 credit crisis \cite{fennchaos}, which has important implications for the robustness of financial markets \cite{may}. Financial institutions are linked both through credit relationships and as a result of holding similar portfolios of assets \cite{allenbabus}. If many assets are correlated and prices fall, this can cause several financial institutions to write down the value of their assets. These write-downs can then impact the credit relationships between different institutions \cite{gai}. The strength of correlations therefore affects the stability of markets.

The rest of this paper is organized as follows.  In Section \ref{sec:data}, we discuss our data.  We then discuss correlations between assets in Section \ref{sec:pca:corr}, PCA and RMT in Section \ref{sec:pca}, the temporal evolution of assets in Section \ref{sec:tempevo}, correlations between individual assets and components in Section \ref{sec:assetcorr}, and individual asset classes in Section \ref{sec:individualassets}. We summarize our results in Section \ref{sec:pcaconclusions}. We enumerate the assets that we considered in Appendix~\ref{sec:assetlist}, provide examples of the changes in the correlations between assets and principal components (PCs) in Appendix~\ref{sec:egcorrs}, and consider the contribution of the assets to these correlations in Appendix~\ref{sec:selfcorr}.

%%%%%%%%%%%%%%%%%%%%%%%%%%%%%%%%%%%%%%%%%%%%%%%%%%%%%%%%%%%%%%%%%%%

\section{Data} \label{sec:data}

We examine time series for $N = 98$ financial products for the period 01/08/1999--01/01/2010.  These products include 25 developed market equity indices, 3 emerging market equity indices, 4 corporate bond indices, 20 government bond indices, 15 currencies, 9 metals, 4 fuel commodities, and 18 other commodities. (We enumerate and describe these assets in Table \ref{tbl:assets} of the appendix.) We include markets from several geographical regions, so many are traded during different hours of the day. For example, stocks included in the Nikkei 225 are traded on the Tokyo Stock Exchange, which operates from midnight to 6 AM GMT, whereas stocks included in the FTSE 100 index are traded on the London Stock Exchange, which operates from 8 AM to 4:30 PM GMT. We use weekly price data to minimize effects that might result from the different trading hours for markets from different time zones.

%%%%%%%%%%%%%%%%%%%%%%%%%%%%%%%%%%%%%%%%%%%%%%%%%%%%%%%%%%%%%%%%%%%

\section{Correlations} \label{sec:pca:corr}

We denote the price of asset $i$ at discrete time $t$ as $p_i(t)$ ($i=1,\ldots,N$) and define a weekly logarithmic return $z_i(t)$ for asset $i$
%between consecutive time steps
as
\begin{equation}
	z_i(t)=\ln\left[\frac{p_i(t)}{p_i(t-1)}\right]\,. \label{eq:pca:returns}
\end{equation}
We define a standardized return as $\hat{z}_i(t)=[z_i(t)-\left\langle z_i \right\rangle]/\sigma(z_i)$, where $\sigma(z_i) = \sqrt{\left\langle z_i^2 \right\rangle-\left\langle z_i \right\rangle^2}$ is the standard deviation of $z_i$ over a time window of $T$ time steps and $\left\langle\cdots\right\rangle$ denotes a time average over the same time window. We then represent the standardized returns as an $N\times T$ matrix $\hat{\mathbf{Z}}$, so the empirical correlation matrix is
\begin{equation}
	\mathbf{R}=\frac{1}{T}\hat{\mathbf{Z}}\hat{\mathbf{Z}}^T\,,
\label{eq:pcacorr}
\end{equation}
which has elements $r(i,j) \in [-1,1]$.  Because we have standardized the time series, the correlation matrix $\mathbf{R}$ of returns $\hat{\mathbf{Z}}$ is equal to the covariance matrix $\mathbf{\Sigma}_{\hat{\mathbf{Z}}}$ of $\hat{\mathbf{Z}}$.

\begin{figure}[htp]
\begin{center}
\includegraphics[width=1\linewidth]{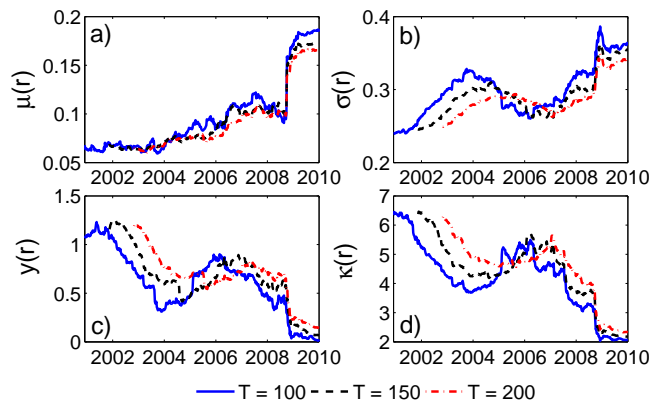}
\caption{(Color online) Comparison of the distribution of correlation coefficients as a function of time for time windows of length $T=100$, $150$, and $200$. (a) Mean correlation $\mu(r)$, (b) standard deviation $\sigma(r)$, (c) skewness $y(r)$, and (d) kurtosis $\kappa(r)$.}
\label{fig:window_comp}
\end{center}
\end{figure}

We create a time-evolving sequence of correlation matrices by rolling the time window of $T$ returns through the full data set. The choice of $T$ is a compromise between overly noisy and overly smoothed correlation coefficients \cite{nyse} and is usually chosen such that $Q=T/N\geq1$. Figure~\ref{fig:window_comp} shows that we identify the same major changes in the distribution of correlation coefficients using different values of $T$; however, some of the features in the correlations are smoothed out for longer windows. In this study, we fix $T=100$ (each window then contains just under two years of data and $Q \doteq 1.02$), and we roll the time window through the data one week at a time. By only shifting the time window by one data point, there is a significant overlap in the data contained in consecutive windows.  However, this approach enables us to track the evolution of the market correlations and to identify time steps at which there were significant changes in the correlations. The choice of $T=100$ results in 475 correlation matrices for the period 1999--2010.

\begin{figure}[htp]
\begin{center}
\includegraphics[width=1\linewidth]{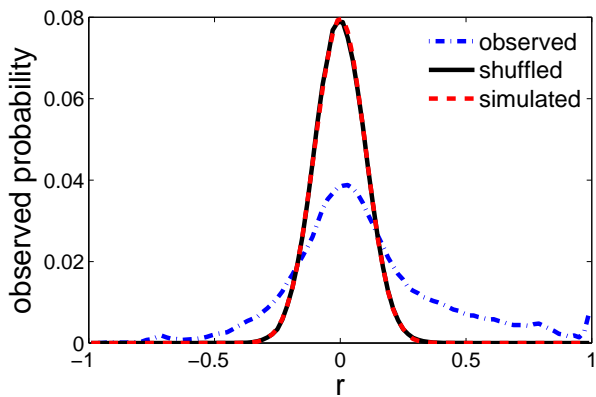}
\caption{(Color online) Distribution of all of the correlation coefficients $r(i,j)$ from every time window for market and random data. The shuffled and simulated data lie almost on top of each other.}
\label{fig:cc_dist}
\end{center}
\end{figure}

In Fig.~\ref{fig:cc_dist}, we show the distribution of all of the empirical correlation coefficients from every time window. To highlight interesting features in the correlations, we compare the distribution to corresponding distributions for simulated random returns and randomly shuffled returns. We generate shuffled data by randomly reordering the full return time series for each asset independently.  This process destroys the temporal correlations between the return time series but preserves the distribution of returns for each series. We then produce correlation matrices for the shuffled returns by rolling a time window of $T$ time steps through the shuffled data and calculating a correlation matrix for each position of the window. We produce simulated data by independently generating $N$ time series of returns (where each series has the same length as the original data) whose elements are drawn from a Gaussian distribution with mean zero and unit variance. We again roll a time window of length $T$ through the data and calculate a correlation matrix for each window \footnote{Return time series for some financial assets display ``volatility clustering'', which means that large-magnitude returns tend to be followed by large-magnitude returns and small-magnitude returns tend to be followed by small-magnitude returns \cite{mandelbrot,hff}. The random return time series that we investigate do not possess this property; moreover, such memory of volatility in individual time series would not affect the results. We consider zero-lag correlations and the zero-lag autocorrelation is necessarily equal to 1 for any time series \cite{tsay}. Given this, the results for an investigation of correlations at zero lag for mutually uncorrelated time series with memory must be the same as the results for mutually uncorrelated time series without memory.}. 

Figure~\ref{fig:cc_dist} illustrates that the distribution of correlation coefficients for the market data is significantly different from the two random distributions as the market data has more large positive and negative correlations. The differences between the distributions demonstrate that there are temporal correlations between returns for financial assets that are incompatible with the null models that we consider, which in turn implies that financial market correlation matrices contain structure that warrants investigation.

%%%%%%%%%%%%%%%%%%%%

\section{Principal Component Analysis and Random Matrix Theory} \label{sec:pca}

We investigate the structure of the correlation matrices using PCA. The aim of PCA is to find the linear transformation $\mathbf{\Omega}$ that maps a set of observed variables $\hat{\mathbf{Z}}$ into a set of uncorrelated variables $\mathbf{Y}$ \cite{pcajolliffe}. We define the $N\times T$ matrix 
\begin{equation}
	\mathbf{Y}=\mathbf{\Omega}\hat{\mathbf{Z}}\,,
\label{eq:pcs}
\end{equation}
where each row $\mathbf{y}_k$ ($k=1,\ldots, N$) corresponds to a PC of $\hat{\mathbf{Z}}$ and the transformation matrix $\mathbf{\Omega}$ has elements $\omega_{ki}$. The first row $\mathbf{\omega}_1$ of $\mathbf{\Omega}$ (which contains the first set of PC coefficients) is chosen such that the first PC $\mathbf{y}_1$ is aligned with the direction of maximal variance in the $N$-dimensional space defined by $\hat{\mathbf{Z}}$. Each subsequent PC accounts for as much of the remaining variance of $\hat{\mathbf{Z}}$ as possible, subject to the constraint that the $\mathbf{\omega}_k$ are mutually orthogonal. We further constrain the vectors $\mathbf{\omega}_k$ such that $\mathbf{\omega}_k\mathbf{\omega}_k^T=1$ for all $k$.

The correlation matrix $\mathbf{R}$ is an $N\times N$ diagonalizable, symmetric matrix that can be written in the form
\begin{equation}
	\mathbf{R}=\frac{1}{T}\mathbf{EDE}^T\,, \label{eq:cdiag}
\end{equation}
where $\mathbf{D}$ is a diagonal matrix of eigenvalues $\beta_k$ and $\mathbf{E}$ is an orthogonal matrix of the corresponding eigenvectors. It is known \cite{pcajolliffe} that the eigenvectors of the correlation matrix correspond to the directions of maximal variance such that $\mathbf{\Omega}=\mathbf{E}^T$, and one finds the PCs via the diagonalization in Eq.~(\ref{eq:cdiag}). The signs of the PCs are arbitrary; if the sign of every coefficient in a component $\mathbf{y}_k$ is reversed, neither the variance of $\mathbf{y}_k$ nor the orthogonality of $\mathbf{\omega}_k$ with respect to each of the other eigenvectors changes.

%%%%%%%%%%%%

%\subsection{Eigenvalues} \label{sec:eigval}

We compare the properties of the market correlation matrices with correlation matrices for random time series. The correlation matrix for $N$ mutually uncorrelated time series of length $T$ with elements drawn from a Gaussian distribution is a Wishart matrix \cite{Laloux_PRL_1999,Plerou_PRL_1999}. In the limit $N\rightarrow\infty$, $T\rightarrow\infty$, and with the constraint that $Q=T/N\geq{1}$, the probability density function $\rho(\gamma)$ of the eigenvalues $\gamma$ of such correlation matrices is given by \cite{sengupta}
\begin{equation}
\rho(\gamma)=\frac{Q}{2\pi\sigma^2(\hat{\mathbf{Z}})}\frac{\sqrt{(\gamma_{+}-\gamma)(\gamma_{-}-\gamma)}}{\gamma}\,, \label{eq:RMT_eig_dist}
\end{equation}
where $\sigma^2(\hat{\mathbf{Z}})$ denotes the variance of the elements of $\hat{\mathbf{Z}}$, and $\gamma_{+}$ and $\gamma_{-}$ are the maximum and minimum eigenvalues of the matrix, which are given by
\begin{equation}
	\gamma_{\pm}=\sigma^2(\hat{\mathbf{Z}})\left(1+\frac{1}{Q}\pm 2\sqrt{\frac{1}{Q}}\right)\,.
\label{eq:RMT_bounds}
\end{equation}
When $Q=1$, the minimum eigenvalue is $\gamma_-=0$, the maximum eigenvalue is $\gamma_+=4\sigma^2(\hat{\mathbf{Z}})$, and the density $\rho(\gamma)$ diverges as $1/\sqrt{\gamma}$ as $\gamma\rightarrow\gamma_- = 0$.  Equations~(\ref{eq:RMT_eig_dist},\ref{eq:RMT_bounds}) are only valid in the limit $N\rightarrow\infty$.  For finite $N$, there is a non-zero probability of finding eigenvalues larger than $\gamma_+$ and smaller than $\gamma_-$. For the returns that we investigate, $\sigma^2(\hat{\mathbf{Z}})=1$ and $\gamma_+\doteq{3.96}$.

\begin{figure}[htp]
\begin{center}
\includegraphics[width=1\linewidth]{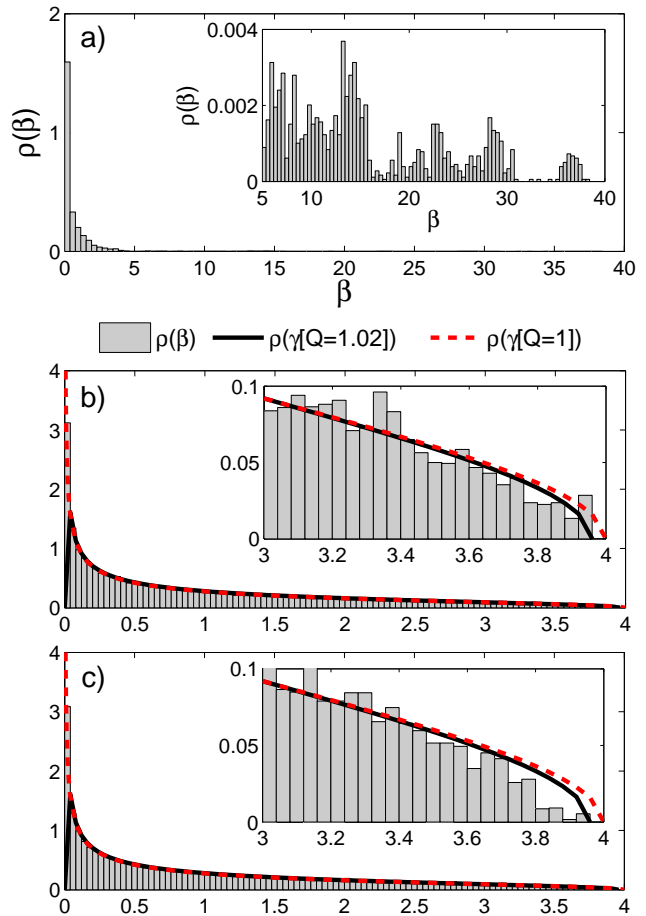}
\caption{(Color online) Distribution of eigenvalues $\rho(\beta)$ of the correlation matrices for all time windows for (a) market, (b) shuffled, and (c) simulated data. The insets show the distributions of the largest eigenvalues. In (b) and (c), we show the eigenvalue probability density functions $\rho(\gamma)$ for random matrices given by Eq.~(\ref{eq:RMT_eig_dist}) for $Q=1.02$ and $Q = 1$.}
\label{fig:eig_dist}
\end{center}
\end{figure}

In Fig.~\ref{fig:eig_dist}, we compare the distribution of all eigenvalues from every time window for market data with the distributions for shuffled and simulated data. In panel (a), we show that the eigenvalue distribution for market correlations differs from that of random matrices: there are many eigenvalues larger than the upper bound $\gamma_+\doteq{3.96}$ predicted by RMT (with several eigenvalues almost 10 times as large as the upper bound). In prior studies of equity markets, the eigenvector corresponding to the largest eigenvalue has been described as a ``market'' component, with roughly equal contributions from each of the $N$ equities studied, and the eigenvectors corresponding to the other eigenvalues larger than $\gamma_+$ have been identified as different market sectors \cite{Laloux_PRL_1999,Plerou_PRL_1999}. In Section~\ref{sec:assetcorr}, we discuss the interpretation of the observed eigenvectors with eigenvalues $\beta>\gamma_+$. For now, we simply note that the deviations of the empirical distribution of eigenvalues from the predictions of RMT again imply that the correlation matrices contain structure that is incompatible with the null models that we consider.

In Figs.~\ref{fig:eig_dist}(b) and (c), we illustrate that the distributions for shuffled and simulated data are very similar and that they agree very well with the analytical distribution given by Eq.~(\ref{eq:RMT_eig_dist}) over most of the range of $\gamma$.  In particular, both distributions have an upper bound close to the theoretical maximum $\gamma_+\doteq{3.96}$. However, for $Q=1.02$ (the value that corresponds to the selected $T$ and $N$), the observed distribution of eigenvalues for random data does not fit the distribution in Eq.~(\ref{eq:RMT_eig_dist}) as $\gamma\rightarrow{0}$.  For both the simulated and shuffled data, we observe a much higher probability density near $\gamma = 0$ than that predicted by RMT. The high probability density near zero is a result of the fact that $T\approx N$. When we simulate eigenvalue distributions for data with $T\gg N$, we observe a much smaller probability density near zero. In Figs.~\ref{fig:eig_dist}(b) and (c), we also show the theoretical distribution for $Q = 1$.  In this case $\rho(\gamma)$ diverges as $\gamma\rightarrow{0}$, which fits the randomly generated distributions reasonably well.

We obtain similar results for the distributions of the elements $\omega_{ki}$ (the PC coefficients) of the eigenvectors of the correlation matrices. Correlation matrices $\mathbf{R}$ are real symmetric matrices, so we compare the eigenvector properties of the matrices $\mathbf{R}$ with those for real symmetric random matrices. Such random matrices display the universal properties of the canonical ensemble of matrices known as the Gaussian orthogonal ensemble (GOE) \cite{Plerou_PRL_1999,Plerou_PRE_2002}. For the GOE, the probability density $\rho(\omega_k)$ of the elements of the $k^\textrm{th}$ eigenvector is a Gaussian distribution with mean zero and unit variance \cite{guhr}. We find that the distributions for shuffled and simulated data closely match a Gaussian distribution, but there are differences between these distributions and the distributions for market correlations. These differences are most pronounced for the first and second PCs. In particular, there are asymmetries in the distributions for market data that are not present in the random distributions.

\section{Temporal Evolution} \label{sec:tempevo}

In prior sections, we studied aggregate results for all time steps and illustrated that the eigenvalues and eigenvectors of financial market correlation matrices suggest that there are correlations that are incompatible with the random null models that we consider. We now investigate the temporal evolution of financial market correlations by investigating changes in the eigenvalues and eigenvectors.

%%%%%%%%%%%%%

\subsection{Proportion of Variance} \label{sec:tempevo:propvar}

We begin by considering the eigenvalues of the correlation matrices. The covariance matrix $\mathbf{\Sigma_Y}$ for the PC matrix $\mathbf{Y}$ can be written as
\begin{equation}
	\mathbf{\Sigma_{Y}}=\frac{1}{T}\mathbf{YY}^T=\frac{1}{T}\mathbf{\Omega}\hat{\mathbf{Z}}\hat{\mathbf{Z}}^T\mathbf{\Omega}^T=\mathbf{D}\,,
\end{equation}
where $\mathbf{D}$ is the diagonal matrix of eigenvalues $\beta$. The total variance of the returns $\hat{\mathbf{Z}}$ for the $N$ assets is then
\begin{equation}
	\sum_{i=1}^N\sigma^2(\hat{\mathbf{z}}_i)=\textrm{tr}(\mathbf{\Sigma}_{\hat{\mathbf{Z}}})=\sum_{i=1}^N\beta_i=\sum_{i=1}^N\sigma^2(\mathbf{y}_i)=\textrm{tr}(\mathbf{D})=N\,,
\label{eq:vsumN}
\end{equation}
where $\mathbf{\Sigma}_{\hat{\mathbf{Z}}}$ is the covariance matrix for $\hat{\mathbf{Z}}$ and $\sigma^2(\hat{\mathbf{z}}_i)=1$ is the variance of the vector $\hat{\mathbf{z}}_i$ of returns for asset $i$. The proportion of the total variance in $\hat{\mathbf{Z}}$ explained by the $k^\textrm{th}$ PC is then
\begin{equation}
	\frac{\sigma^2(\mathbf{y}_k)}{\sum_{i=1}^N\sigma^2(\mathbf{z}_i)}=\frac{\beta_k}{\beta_1+\ldots+\beta_N}=\frac{\beta_k}{N}\,. \label{eq:eigfrac}
\end{equation}
In words, the ratio of the $k^\textrm{th}$ largest eigenvalue $\beta_k$ to the number of assets $N$ is equal to the proportion of the variance from the $k^\textrm{th}$ PC.

\begin{figure}[htp]
\begin{center}
\includegraphics[width=1\linewidth]{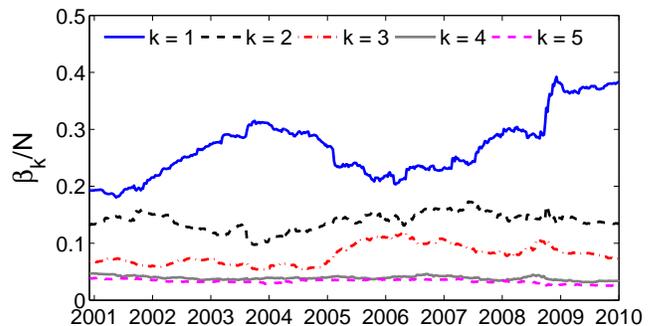}
\caption{(Color online) Fraction of the variance in $\hat{\mathbf{Z}}$ explained by the first five PCs versus time. The top curve shows the variance explained by the first PC, the next curve shows the variance explained by the second PC, and so on. The horizontal axis shows the year of the last data point in each time window.}
\label{fig:eig_t}
\end{center}
\end{figure}

In Fig.~\ref{fig:eig_t}, we show as a function of time the fraction of the variance $\beta_k/N$ due to the first five PCs ($k=1,\ldots,5$). From 2001 to 2004, the fraction of the variance explained by the first PC increased.  Between 2004 and 2006, it decreased before gradually increasing again. In particular, a sharp rise occurred when the week including 09/15/2008 entered the rolling time window. This was the day that Lehman Brothers filed for bankruptcy and Merrill Lynch agreed to be taken over by Bank of America. Both events represented major shocks to the financial system \cite{fennchaos}. The variance explained by the first PC peaked as the week ending 12/05/2008 entered the rolling window---this was the week during which the National Bureau of Economic Research officially declared that the U.S. was in a recession---at which point it accounted for nearly 40\% of the variance in $\hat{\mathbf{Z}}$.

The large variance in market returns explained by a single component implies that there is a large amount of common variation in financial markets. The increase in the variance accounted by the first PC between 2001 and 2010 also suggests that markets have become more correlated in recent years. In particular, the significant rise in the variance of the first PC following the collapse of Lehman Brothers demonstrates that markets became more correlated during the period of crisis following the failure of this major bank.

Although the changes in the variance from other high PCs are smaller than those for the first PC, from 2001 until the collapse of Lehman Brothers, the variance explained by the second and third PCs appears to be negatively correlated with the variance explained by the first PC. This is to be expected because the total variance is constrained to sum to $N$ [see Eq.~(\ref{eq:vsumN})], so when the first PC accounts for a higher proportion, less remains to be explained by the other components. Following Lehman's bankruptcy, there is a sharp increase in the variance explained by the first PC, for which the small decreases in the variances of the next four PCs do not account. Instead, the increase in the variance of the first PC arises from small decreases in the variance of many other PCs. From 09/19/2008 to 12/05/2008, the variance explained by the first PC increased by 10\%. Over the same period, the variance explained by only four other PCs increased (with a maximum increase of 0.02\% for the ninth PC). The variances due to all other PCs decreased. For example, the second, third, fourth, and fifth PCs fell by 0.6\%, 1.3\%, 0.7\%, and 0.5\%, respectively. The combined decrease in these PCs offset the sharp rise in the first PC.

It is also instructive to consider the combined variance explained by the first few PCs. In 2001, the first twelve PCs accounted for approximately 65\% of the variance of market returns. By 2010, however, the first five PCs accounted for the same proportion. The fact that only a few components accounted for such a large proportion of the variance implies that there is a lot of common variation in the return time series of many financial assets and highlights the close ties between different markets. It also suggests that market correlations can be characterized by much fewer than $N$ components.

%%%%%%%%%%%%%

\subsection{Significant Principal Component Coefficients} \label{sec:sigpcc}

An increase in the variance for which a PC accounts might be the result of increases in the correlations among only a few assets (which then have large PC coefficients) or a market-wide effect in which many assets begin to make significant contributions to the component. This is an important distinction, because the two types of changes have very different financial implications. For example, in optimal portfolio selection, it becomes much more difficult to reduce risk by diversifying across different asset classes when correlations between all assets increase.  In contrast, increases in correlations within an asset class that are not accompanied by increases in correlations between asset classes have a less significant impact on diversification. A market-wide increase in correlations might also imply a change in the global macroeconomic environment. If many assets are correlated, this suggests that the same macroeconomic force is driving different markets.

We use the \emph{inverse participation ratio} (IPR) \cite{guhr,Plerou_PRE_2002} to investigate temporal changes in the number of assets that make significant contributions to each component. The IPR $I_k$ of the $k^\textrm{th}$ PC $\mathbf{\omega}_k$ is defined as
\begin{equation}
	I_k=\sum_{i=1}^N[\omega_{ki}]^4\,.
\end{equation}
The IPR quantifies the reciprocal of the number of elements that make a significant contribution to each eigenvector.  Two limiting cases help one to understand it: (1) an eigenvector with identical contributions $\omega_{ki}=1/\sqrt{N}$ from all $N$ assets has $I_k=1/N$; and (2) an eigenvector with a single component $\omega_{ki}=1$ and remaining components equal to zero has $I_k=1$. We also define a \textit{participation ratio} (PR) as $1/I_k$. A large PR for a PC indicates that many assets contribute to it.

\begin{figure}[htp]
\begin{center}
\includegraphics[width=1\linewidth]{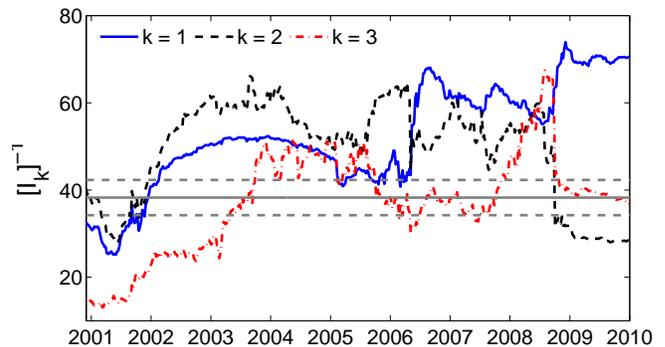}\\
\caption{(Color online) Participation ratio $\left[I_k\right]^{-1}$ as a function of time for the three PCs with the largest variance ($k=1, 2, 3$). The horizontal solid line shows the PR (averaged over 100,000 simulations) of the first PC for randomized returns, and the horizontal dashed lines show one standard deviation above and below the mean.}
\label{fig:ipr}
\end{center}
\end{figure}

In Fig.~\ref{fig:ipr}, we show as a function of time the PR of the first three PCs. The PR of the first PC increased from 2001 to 2010, and there were sharp increases when the weeks ending 05/12/2006 and 09/19/2008 entered the rolling time window. The second increase coincided with the market turmoil that followed the collapse of Lehman Brothers and occurred at the same time as a significant increase in the variance that was explained by the first PC (see Fig.~\ref{fig:eig_t}). The first increase coincided with surging metal prices. During the week ending 05/12/2006, the price of gold rose to a 25-year high, reaching over \$700 per ounce, and the prices of several other metals also rose to record levels.  Platinum and copper reached all time highs, aluminum hit an 18-year peak, and silver prices rose to their highest levels since February 1998.  During the same week, corporate bond prices reached a 2-year high and the prices of emerging market equities reached record levels. Although these events coincided with a significant increase in the PR of the first PC, this increase was not accompanied by a sharp rise in the variance explained by this component.

The sharp rise in the PR of the first PC following the collapse of Lehman Brothers implies that many assets were highly correlated during the ensuing financial crisis.  Based on the value of the PR, over 70\% of the studied assets contributed significantly to the first PC. To test the significance of the PR of the first PC, we compare it to the corresponding PR for random returns. Figure~\ref{fig:ipr} illustrates that between 2006 and 2010, the PR of the observed returns was significantly larger than that expected for random returns, which emphasizes the large number of different assets that were correlated during this period.

The temporal evolutions of the PRs of the higher components are rather different. For example, from 2001 to 2003, the PR of the second PC doubled; it then fluctuated around the same level until the collapse of Lehman Brothers, at which point it decreased sharply. Similarly, the PR of the third PC increased from 2001 until Lehman's collapse, and then it also fell sharply. This suggests that, following Lehman's bankruptcy, the first PC influenced many assets at the expense of higher components. The dominance of a single PC again implies that there is a large amount of common variance in asset returns. It also suggests that the key market correlations can be described using only a few PCs.

From 2001 to 2002, the PR of each of the first three PCs was below the value expected for random returns for all but 9 weeks. For random returns, there are only small differences in the PRs for the different PCs. For example, using 100,000 simulations we find that the mean PR of the first, second, and third PCs are 38.3, 37.7, and 37.3, respectively. The standard deviations are 4.0, 4.1, and 4.2, respectively. This implies that eigenvectors for correlation matrices of random returns are {\it extended} as many different assets contribute to them \cite{Plerou_PRE_2002}. In contrast, from 2001 to 2002, the eigenvectors for the correlation matrices for market data are {\it localized} and have fewer assets contributing to them than expected for correlation matrices for random uncorrelated returns.

Inspection of the first three eigenvectors over this period suggests that they correspond to bonds, equities, and currencies, and the PRs of the first three PCs support this observation. We study 24 bond indices (government and corporate), 28 equity indices, and 15 currencies. At the beginning of 2001 the PRs of these first three PCs were 32.6, 40.3, and 14.9, respectively. As expected, these PRs are larger than the number of bonds, equities, and currencies that we study. The different asset classes have some common variance with other types of assets, which inflates the PR. Taking this effect into account, the PRs are consistent with localized eigenvectors that represent specific asset classes.

From 2001 to 2010, the first PC changed from a localized state in which only bonds contributed significantly to an extended state in which nearly all of the assets that we consider contributed. We discuss this in more detail in Section~\ref{sec:assetcorr}. The large PR during the post-Lehman period and the small PR from 2001 to 2002 are both indicative of correlations that are incompatible with uncorrelated asset returns.

%%%%%%%%%%

\subsection{Number of Significant Components} \label{sec:nosigcomp}

We now attempt to determine how many PCs are needed to describe the primary market correlations. PCA is widely used to produce lower-dimensional representations of multivariate data by retaining a few ``significant'' components and discarding all other components \cite{pcajolliffe}. Many heuristic methods have been proposed for determining the number of significant PCs, but there is no widespread agreement on an optimal approach \cite{nopc}.

\begin{figure}[htp]
\begin{center}
\includegraphics[width=1\linewidth]{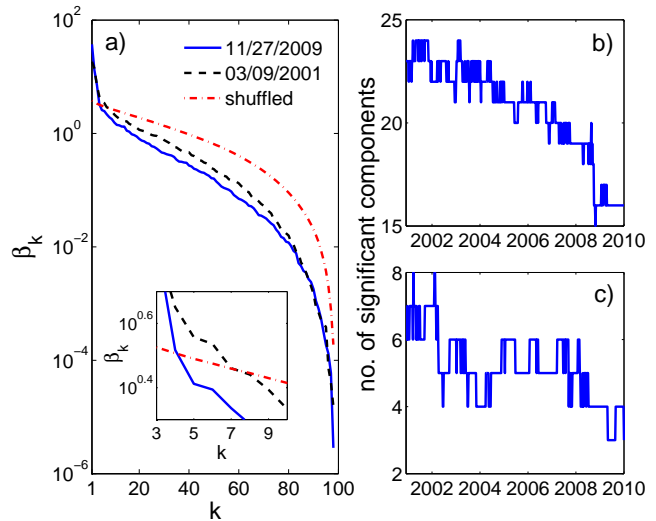}
\caption{(Color online) Panel (a) shows a scree plot, which gives the magnitude of the PC eigenvalues as a function of the eigenvalue index, where the eigenvalues are sorted such that $\beta_1 \geq \beta_2 \geq \ldots \geq \beta_N$. We show curves for random correlation matrices and for correlation matrices for time windows ending on 03/09/2001 and 11/27/2009. The inset zooms in on the region in which the two example curves for observed data cross the curve for random data. Panels (b) and (c) show the number of significant components as a function of time determined using (b) the Kaiser-Guttman criterion and (c) by comparing the scree plots of the observed and random data.
}
\label{fig:no_comp}
\end{center}
\end{figure}

We apply two techniques to find the number of significant components. The first is the Kaiser-Guttman criterion \cite{kaiser}, which assumes that a PC is significant if its eigenvalue $\beta>1/N$. Any component that satisfies this criterion accounts for more than a fraction $(1/N)$ of the variance of the system. It is considered significant because it is assumed to summarize more information than any single original variable. The second approach is to compare the observed eigenvalues to the eigenvalues for random data and can be understood by considering the scree plot in Fig.~\ref{fig:no_comp}(a). A scree plot shows the magnitudes of the eigenvalues as a function of the eigenvalue index, where the eigenvalues are sorted such that $\beta_1 \geq \beta_2 \geq \ldots \geq \beta_N$.  The leftmost data point in a scree plot indicates the magnitude of the largest eigenvalue, and the rightmost data point indicates the magnitude of the smallest eigenvalue. The number of significant PCs is given by the number of eigenvalues in the scree plot for which the eigenvalue for the observed data is larger than the corresponding eigenvalue for random data.  For example, in Fig.~\ref{fig:no_comp}(a), there are 3 significant eigenvalues for 11/27/2009 and 6 significant eigenvalues for 03/09/2001.

Figures~\ref{fig:no_comp}(b) and (c) illustrate that there are large differences in the number of significant components identified using the two techniques, though both agree that the number decreased between 2001 and 2010. The discrepancies in the results suggest that one cannot reliably determine the exact number of significant PCs using these two methods.  Nevertheless, the similar trends obtained using the two techniques provide evidence that the number of significant components decreased from 2001 to 2010. This again implies that markets have become more correlated in recent years. Both methods also agree that the number of significant components is much lower than the number of assets that we considered. Therefore, although one cannot determine precisely the number of significant components using the methods described in this section, our results nonetheless suggest that market correlations can be characterized by much fewer than $N$ components.

%%%%%%%%%%%%

\section{Correlations Betweens Assets and Components} \label{sec:assetcorr}

We now return to the question of the interpretation of the eigenvectors with eigenvalues $\beta$ larger than the upper bound $\gamma_+$ predicted by RMT. To do this, we investigate the correlations $r(\hat{\mathbf{z}}_i,\mathbf{y}_k)$ between the asset return time series $\hat{\mathbf{z}}_i$ and the PCs $\mathbf{y}_k$. These correlations are closely related to the PC coefficients, which represent the weighting of each asset on the PCs.  However, because the correlations $r(\hat{\mathbf{z}}_i,\mathbf{y}_k)$ are confined to the interval $[-1,1]$, they are easier to interpret than the PC coefficients. We use the correlations $r(\hat{\mathbf{z}}_i,\mathbf{y}_k)$ to measure the strengths of the asset-PC relationships and to determine which assets contribute to each PC. In doing this, we also determine the number of PCs that need to be retained to describe the main features of the correlation matrices.

We write the covariance matrix of the return $\hat{\mathbf{Z}}$ with PCs $\mathbf{Y}$ as
\begin{equation}
	\mathbf{\Sigma}_{\mathbf{Y}\hat{\mathbf{Z}}}=\frac{1}{T}\mathbf{Y}{\hat{\mathbf{Z}}}^T=\frac{1}{T}\mathbf{\Omega}\hat{\mathbf{Z}}{\hat{\mathbf{Z}}}^T=\mathbf{\Omega\Omega}^T\mathbf{D\Omega}=\mathbf{D\Omega}\,.
\end{equation}
This implies that the covariance of the returns of asset $i$ and the
$k^{\textrm{th}}$ PC is given by $\Sigma(\hat{\mathbf{z}}_i,\mathbf{y}_k)=\omega_{ki}\beta_k$.  Additionally, the correlation $r(\hat{\mathbf{z}}_i,\mathbf{y}_k)$ is given by
\begin{equation}	r(\hat{\mathbf{z}}_i,\mathbf{y}_k)=\frac{\omega_{ki}\beta_{k}}{\sigma(\hat{\mathbf{z}}_i)\sigma(\mathbf{y}_k)}=\omega_{ki}\sqrt{\beta_k}\,,
\end{equation}
where $\sigma(\hat{\mathbf{z}}_i)=1$ is the standard deviation of $\hat{\mathbf{z}}_i$ over $T$ returns and $\sigma(\mathbf{y}_k)=\sqrt{\beta_k}$. The correlations between the PCs and the original variables are therefore equal to the PC coefficients scaled by the appropriate eigenvalue. The signs of the PC coefficients are arbitrary, so the signs of the PCs and the signs of the correlations $r(\hat{\mathbf{z}}_i,\mathbf{y}_k)$ are also arbitrary. To avoid having to choose a sign for each correlation coefficient, we consider absolute correlations $\vert r(\hat{\mathbf{z}}_i,\mathbf{y}_k)\vert$.  Although we can then no longer tell whether an asset is positively or negatively correlated with a PC, this step is reasonable because we are interested only in determining which assets contribute to each component.

\subsection{Assets Correlated with Each Component}
\label{sec:assetpccorr}

\begin{figure*}[htp]
\begin{center}
\includegraphics[width=1\linewidth]{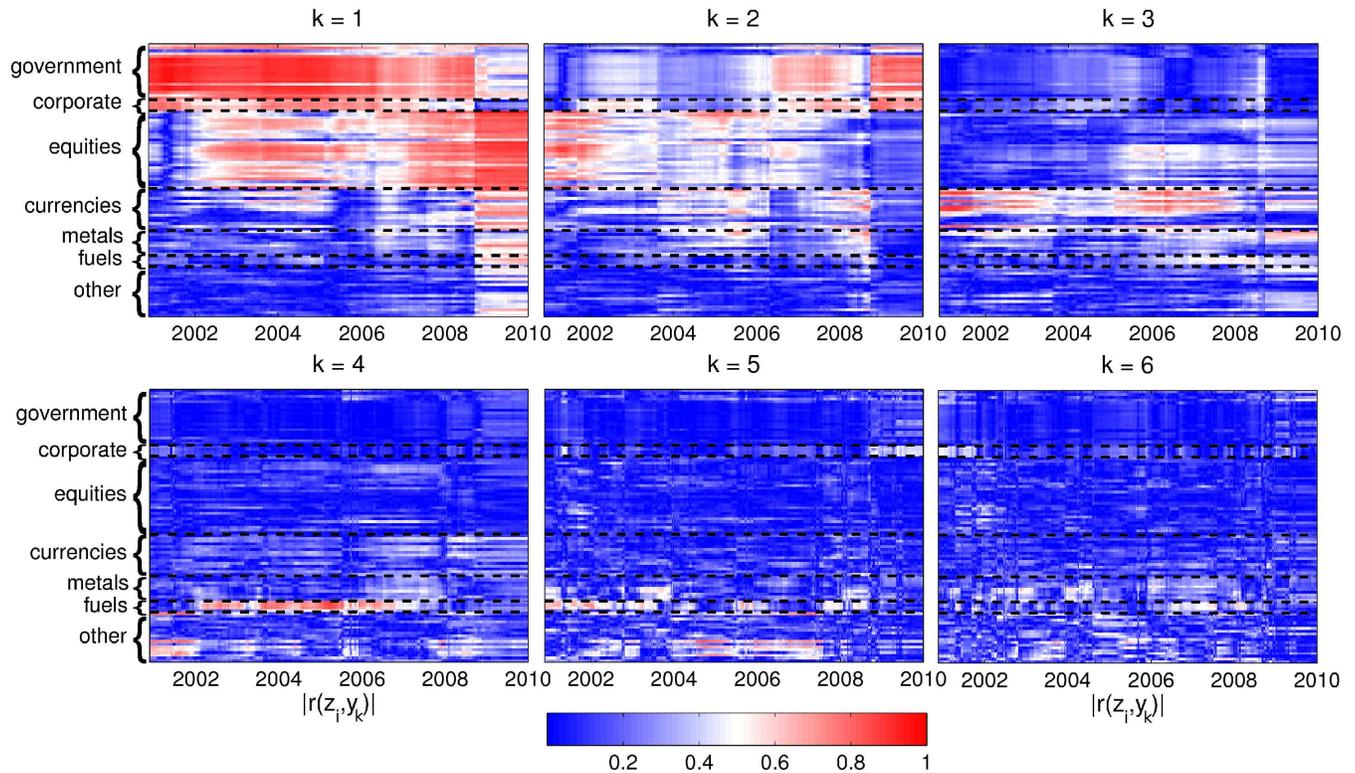}
\caption{(Color online) The absolute correlation $\vert r(\hat{\mathbf{z}}_i,\mathbf{y}_k)\vert$ between each asset and the first six PCs ($k=1,\ldots,6$) as a function of time. Each point on the horizontal axis represents a single time window, and each position along the vertical axis represents an asset.}
\label{fig:sig_corr}
\end{center}
\end{figure*}

In Fig.~\ref{fig:sig_corr}, we show the variation through time of the correlation of every asset with each of the first six PCs. (In Appendix~\ref{sec:egcorrs}, we show example plots of $\vert r(\hat{\mathbf{z}}_i,\mathbf{y}_k)\vert$ as a function of time for specific assets.) This figure highlights that there are significantly fewer high correlations for the components with larger $k$. For example, many of the correlations in the first PC are greater than $0.8$, but the correlations between the asset returns and the sixth PC rarely exceed $0.5$. As one considers increasingly higher components, the maximum correlation decreases until all correlations are less than 0.2 for the highest components. The low correlations between the asset return time series and the higher PCs implies that much of the key structure from the correlation matrices is contained in the first few PCs. Because the assets contribute to the PCs, some of the correlation $\vert r(\hat{\mathbf{z}}_i,\mathbf{y}_k)\vert$ between the $i^\textrm{th}$ asset and the $k^\textrm{th}$ PC is attributable to asset $i$. We discuss the effect of these ``self-correlations'' in Appendix~\ref{sec:selfcorr}.
%Based on the correlations shown in Fig.~\ref{fig:sig_corr}, it appears that the first five PCs describe the main features of the correlations for the assets that we studied.

Figure~\ref{fig:sig_corr} demonstrates the changing correlations between the different asset classes. From 2001 to 2002, all of the corporate and government bonds (except Japanese government bonds) were strongly correlated with the first PC. Over the same period, most of the equity indices were strongly correlated with the second PC and most of the currencies were strongly correlated with the third PC; six grain commodities (soybean, soybean meal, soybean oil, corn, wheat, and oats) were strongly correlated with the fourth PC; and fuel commodities were strongly correlated with the fifth PC. Therefore, over this period, each of the first five PCs corresponded to a specific market, and the separation into components gave low correlations between different assets classes. During 2002, however, these relationships began to break down as bonds and equities both became strongly correlated with the first PC and both types of asset had a correlation of approximately 0.5 with the second PC. The strong correlation of both bonds and equities with the same PCs marked the start of a period during which the coupling between asset classes increased and different markets became more closely related.

We found three major changes in the correlations between 2002 and 2009; these are most clearly seen by examining the second PC in Fig.~\ref{fig:sig_corr}. The first change corresponded to a local peak in corporate bond prices; the second change corresponded to surging metal prices; and the third (and most striking) change occurred following the collapse of Lehman Brothers. After they declared bankruptcy, the first PC became strongly correlated with nearly all assets---including equities, currencies, metals, fuels, other commodities, and some government bonds. The major exceptions were corporate bonds and (to a lesser extent) government bonds, but both sets of bonds were strongly correlated with the second PC. During this period, only a few assets were strongly correlated with the third PC; these included EURUSD, CHFUSD, gold, silver, and platinum.  Additionally, very few assets were strongly correlated with the higher PCs. The strong correlations between the majority of the studied assets and the first PC following Lehman Brothers' collapse further demonstrates the strength of market correlations during this crisis period and highlights the common behavior of nearly all markets. It also suggests that many different markets are being driven by the same macroeconomic forces.

Figure~\ref{fig:sig_corr} also illustrates that for a system in which the first few PCs account for a significant proportion of the variance, a consideration of the correlations between these components and the original variables provides a parsimonious framework to uncover the key relationships in the system. Instead of having to identify important correlations in a matrix with $\frac{1}{2}N(N-1)$ elements, one only needs to consider correlations between the $N$ variables and the first few PCs, which reduces the number of correlations to consider by a factor of $N$. Figure~\ref{fig:sig_corr} demonstrates that this method uncovers the changing relationships between the different asset classes and highlights assets, such as Japanese government bonds, whose behavior was unusual.  As we have discussed above, this approach also uncovers notable changes that occurred in markets as well as the assets that were significantly affected by these changes.

\subsection{Financial Factor Models}

Several models have been proposed that attempt to explain return time series using linear combinations of one or more financial market factors \cite{clm,tsay}. Some of these models are closely related to PCA, so we discuss them briefly. One of the most widely studied {\it factor models} is the Capital Asset Pricing Model (CAPM) \cite{capm1,capm2}, which relates the return of equities to the returns of a single factor--the ``market portfolio''--which is usually taken as the return of a market-wide index like the S\&P 500 \cite{clm}. Empirical evidence indicates, however, that the CAPM does not explain the behavior of all asset returns \cite{clm}. This implies that other factors might be needed to explain fully return time series, which has led to the development of models with multiple factors. One general model is the Arbitrage Pricing Theory (APT) \cite{apt}, which provides an approximate model of expected asset returns using an unknown number of unidentified factors. The problem then becomes to identify the factors.

The approaches for identifying factors fall into two basic categories: statistical and theoretical. The theoretical models are based on specifying macroeconomic variables (such as gross national product or changes in bond yields \cite{crr}) or firm-specific variables (such as market capitalization \cite{tsay}) as factors. Relationships between the variables and the return time series are then often determined using linear regression. Statistical methods make no assumptions about which variables correspond to which factors and instead identify the factors directly from the return time series. Two commonly used statistical methods are {\it factor analysis} and PCA \cite{clm,tsay}.

Many prior studies that use PCA to find factors focus on equity markets \cite{clm,tsay}. In this approach, the return time series are used to construct portfolios that represent factors. The PCs define different portfolios in which the weights of the assets are based on the PC coefficients. One of the issues in defining factors in this way is that the factors strongly depend on the time period that the return time series cover \cite{clm}.

In contrast to many factor models, we consider different types of asset instead of focusing on a particular asset class. This does not prevent us from using PCA to identify portfolios of assets that represent the factors. However, as we showed in Fig.~\ref{fig:sig_corr} and discussed in Section~\ref{sec:assetpccorr}, we found significant variations in the assets that contributed to each PC during different time periods. For example, from 2001--2002, the first PC corresponded to bonds, whereas a wide range of different assets contributed to this component following the collapse of Lehman Brothers. These variations would therefore result in significant changes in the portfolios corresponding to each factor, implying that the PCs do not represent the same factors through time. Consequently, there is no simple interpretation of the PCs as specific financial factors.

%%%%%%%

\section{Individual asset classes}
\label{sec:individualassets}

Finally, we repeat some of our analysis using correlation matrices that only include similar types of assets. For some classes, we possess only a few time series, so we combine asset classes. We consider equities, currencies, bonds (government and corporate), and commodities (which includes all assets categorized as metals, fuels, and commodities in Table~\ref{tbl:assets}).

\subsection{Fraction of Variance for Asset Classes}

\begin{figure}[htp]
\begin{center}
\includegraphics[width=1\linewidth]{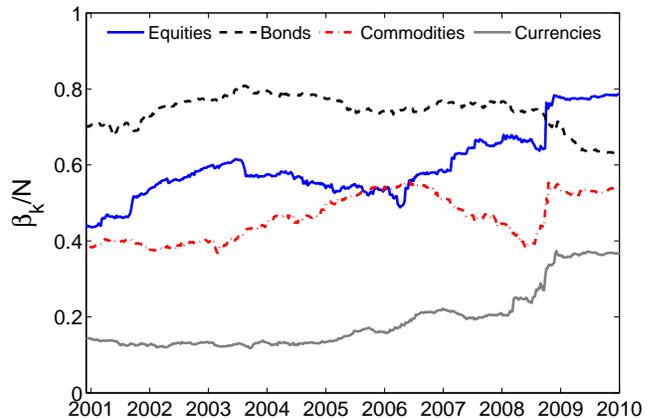}
\caption{(Color online) Fraction of the variance in the returns $\hat{\mathbf{Z}}$ for individual asset classes explained by the first PC versus time. The horizontal axis shows the year of the last data point in each time window.}
\label{fig:eig_t_assets}
\end{center}
\end{figure}

In Fig.~\ref{fig:eig_t_assets}, we show as a function of time the fraction of the variance explained by the first PC for correlation matrices that only include one type of asset. The changes in the variances for equities, commodities, and currencies are similar to those that we observed in Fig.~\ref{fig:eig_t} for all assets. In particular, the variance explained by the first PC increased from 2001 to 2010, and there was a sharp rise in September 2008 following the collapse of Lehman Brothers. In contrast, the variance in bond returns for which the first PC accounted decreased over the same period. From 2001 to 2010, the variance explained by the first five PCs increased from 68.6\% to 89.6\% for equities, from 44.8\% to 60.3\% for commodities, and from 76.1\% to 82.1\% for currencies, but it decreased from 92.4\% to 86.7\% for bonds.

The decrease in the variance of bond returns explained by the first PC following the collapse of Lehman Brothers implies that bonds became less correlated during this period. To explain this change, we consider the correlation between government bonds for the following European countries: Finland, Ireland, Greece, the Netherlands, France, Austria, Belgium, Portugal, Spain, Italy, and Denmark. All of these countries except for Denmark use the Euro, and the value of the Danish krone is pegged to the Euro \footnote{The peg between the Danish krone and the Euro means that the value of the krone is matched to the value of the Euro.}. For the period 01/08/1999--09/12/2008, the mean correlation between these bonds was 0.98 and the standard deviation was 0.01. The high correlation implies that the return time series for the different countries' bonds were very similar prior to the crisis. For the period 09/19/2008--01/01/2010, however, the mean correlation fell to 0.83 and the standard deviation was 0.11. During the crisis, financial uncertainty increased and market participants became more concerned about the higher default risk for particular countries \cite{bonds_imf}. These concerns were heightened by the downgrading of the sovereign debt rating of Greece, Spain, and Portugal, and a flight to safer bonds, such as German bunds, increased variations in the behavior of different bonds \cite{bonds_imf}. The increased variations resulted in lower correlations between bond returns and a reduction in the variance explained by the first PC.

\subsection{Correlations Between Assets and PCs}

We also calculate the correlations between assets and PCs for each asset class to determine which assets contributed to each PC. For equities, nearly all of the indices were strongly correlated with the first PC over the full period. The exceptions were the Nikkei, New Zealand All Ordinaries Index, Austrian Traded Index, and Athens General Index. However, all of these indices were strongly correlated with the first PC after Lehman's bankruptcy. The strong contribution of nearly all equities to the first PC is consistent with prior studies in which this eigenvector was identified as a market factor that affected all stocks \cite{Laloux_PRL_1999,Plerou_PRL_1999,Plerou_PRE_2002}. During some time periods, the second PC corresponded to a group of peripheral Euro-zone countries (including Switzerland, Portugal, Belgium, Ireland, and Austria), whereas it corresponded to emerging market equities during other periods. Similarly, the third PC was correlated with different groups of indices during different time periods. The higher PCs tended to be strongly correlated with single indices, which implies that they represented country-specific factors. There were frequent changes in the indices correlated with the higher PCs. This is the result of changes in the variance of the index returns for the different assets, which affect the ordering of the PCs.

From 2001 to 2010 all of the bond indices except Japanese government bonds were strongly correlated with the first PC. During most of this period the second PC was correlated with corporate bonds. The assets that contributed to the third and fourth PCs changed through time. From 2001--2002 and from 2004--2010, the third PC corresponded to New Zealand and Australian government bonds and the fourth PC corresponded to Japanese government bonds. From 2002--2004, Japanese bonds were strongly correlated with the third PC and New Zealand and Australian bonds were strongly correlated with the fourth PC. The localization of these eigenvectors implies that the PCs represented asset-specific factors, and the change in the identity of the two PCs is again the result of the changing variances of the return time series. As with equities, the higher PCs tended to be correlated with single assets.

The correlations between assets and PCs for commodities and currencies were similar to those shown in Fig.~\ref{fig:sig_corr} for the correlation matrices of all assets. From 2001--2002 the first few PCs for the commodities were correlated with particular types of assets. For example, grain commodities were correlated with the first PC and heating oil, crude oil, and base metals (copper, aluminum, lead, nickel, and tin) were correlated with the second PC. By 2010, however, nearly all of the commodities (with the exception of orange juice, lumber, lean hogs, and pork bellies) were strongly correlated with the first PC, and few assets were strongly correlated with any of the other PCs. Similarly, during 2010, all of the currencies were strongly correlated with the first PC, with the exception of the ``safe haven'' currencies JPY and CHF \cite{safehaven}.

For all of the asset classes, the low correlations between the asset return time series and the higher PCs again indicate that much of the key structure from the correlation matrices is contained in the first few PCs.

%%%%%%%

\section{Summary} \label{sec:pcaconclusions}

We used principal component analysis to investigate the evolving correlation structure of financial markets and to study common features of different markets. We found that the percentage of the variance in market returns explained by the first principal component steadily increased starting in 2006, and that there was a sharp rise following the 2008 collapse of Lehman Brothers. We also found that the number of significant components decreased and that the number of assets making significant contributions to the first PC increased over this period. The strength of the correlations across asset classes following Lehman's bankruptcy suggests that many different markets were being driven by the same macroeconomic forces during the financial crisis. We also showed, however, that both the proportion of the variance explained by the first principal component and the participation ratio of this component increased starting in 2006, which implies that pairwise correlations between a variety of different assets increased for several years before the crisis. It is conceivable that the steady increase in correlations from 2006 might be associated with the growing internationalization of financial markets over this period \cite{lane}, though more research would be necessary to support such a conclusion.

We also investigated the time-evolving relationships between the different assets by investigating the correlations between the asset price time series and the first few PCs. From 2001 to 2002, each of the first five PCs corresponded to a specific market.  However, after 2002, these relationships broke down; by 2010, nearly all of the assets that we studied were significantly correlated with the first PC. We observed similar behavior for correlation matrices of individual asset classes.

%Improvements in information technology have enabled the rapid dissemination of market information around the world and have meant that more market participants now receive a broader range of market data and signals. Before these advances, market participants might have only received news that related directly to the market in which they traded; now, however, news is aggregated and easily accessible so that traders often respond to data from different markets. These changes have potentially contributed to the increased correlations witnessed in markets during recent years.

%%%%%%%%%%

\section*{Acknowledgments}

We thank Mark Austin, David Bloom, Martin Gould, Sam Howison, Paul Mackel, and Craig Veysey for useful discussions, and we acknowledge HSBC bank for providing the data. NSJ acknowledges support from the BBSRC and EPSRC and the grants EP/I005986/1, EP/H046917/1, EP/I005765/1. DJF acknowledges a CASE award from the EPSRC and HSBC bank. MAP acknowledges a research award (\#220020177) from the James S. McDonnell Foundation.

%%%%%%

\appendix

\section{List of Assets}
\label{sec:assetlist}

In Table~\ref{tbl:assets}, we provide details of all of the assets that we study. We selected the assets so that the data includes price time series for all of the major markets. The number of assets $N$ that we include is limited by the constraint that $Q=T/N\geq1$ and the fact that the correlation coefficients are too smoothed if $T$ is too large. Because of these constraints, we use indices for some markets, instead of individual assets, in order to obtain an aggregate view of the market. For all of the commodities, we use futures contracts because commodities are most widely traded in the futures market.

\begin{center}
\begin{longtable*}{p{2.1cm} p{2.2cm} p{5.2cm} p{2.2cm} p{2.2cm} p{3.1cm}}
\caption{\label{tbl:assets}Enumeration and details of all of the assets that we study.}\\

%\begin{tabular}
\hline
\textbf{Ticker} & \textbf{Asset class} & \textbf{Description} & \textbf{Ticker} & \textbf{Asset class} & \textbf{Description} \\ \hline\hline
\endfirsthead

\hline
\textbf{Ticker} & \textbf{Asset class} & \textbf{Description} & \textbf{Ticker} & \textbf{Asset class} & \textbf{Description} \\ \hline\hline
\endhead

\hline \multicolumn{6}{r}{{continued on next page}} \\ \hline
\endfoot

\hline\hline
\endlastfoot

AEX	&	Equities	&	Netherlands AEX Index & CADUSD	&	Currencies	&	Canadian dollar	\\
AS30	&	Equities	&	Australian All Ordinaries Index & CHFUSD	&	Currencies	&	Swiss franc	\\
ASE	&	Equities	&	Athens General Index & CZKUSD	&	Currencies	&	Czech koruna	\\
ATX	&	Equities	&	Austrian Traded Index & EURUSD	&	Currencies	&	Euro	\\
BEL20	&	Equities	&	Belgium BEL 20 Index & GBPUSD	&	Currencies	&	Pounds sterling	\\
BVLX	&	Equities	&	Portugal PSI General Index & IDRUSD	&	Currencies	&	Indonesian rupiah	\\
CAC	&	Equities	&	France CAC 40 Index & JPYUSD	&	Currencies	&	Japanese yen	\\
DAX	&	Equities	&	Germany DAX Index & KRWUSD	&	Currencies	&	Korean won	\\
FTSEMIB	&	Equities	&	Italy FTSE MIB Index & MXNUSD	&	Currencies	&	Mexican peso	\\
HEX	&	Equities	&	Helsinki SX General Index & NOKUSD	&	Currencies	&	Norwegian krone	\\
HSI	&	Equities	&	Hong Kong Hang Seng Index & NZDUSD	&	Currencies	&	New Zealand dollar	\\
IBEX	&	Equities	&	Spain IBEX 35 Index & PHPUSD	&	Currencies	&	Philippines peso	\\
INDU	&	Equities	&	Dow Jones Industrial Average Index & SEKUSD	&	Currencies	&	Swedish krona	\\
ISEQ	&	Equities	&	Irish Overall Index & ZARUSD	&	Currencies	&	South African rand	\\
KFX	&	Equities	&	OMX Copenhagen 20 Index & HG1	&	Metals	&	Copper	\\
NDX	&	Equities	&	NASDAQ 100 Index & LA1	&	Metals	&	Aluminum	\\
NKY	&	Equities	&	Nikkei 225 Index & LL1	&	Metals	&	Lead	\\
NZSE	&	Equities	&	New Zealand All Ordinaries Index & LN1	&	Metals	&	Nickel	\\
OBX	&	Equities	&	Norway OBX Stock Index & LT1	&	Metals	&	Tin	\\
OMX	&	Equities	&	OMX Stockholm 30 Index & XAG	&	Metals	&	Silver	\\
RTY	&	Equities	&	Russell 2000 Index & XAU	&	Metals	&	Gold	\\
SMI	&	Equities	&	Swiss Market Index & XPD	&	Metals	&	Palladium	\\
SPTSX	&	Equities	&	S\&P/Toronto SX Composite Index & XPT	&	Metals	&	Platinum	\\
SPX	&	Equities	&	Standard and Poor's 500 & CL1	&	Fuels	&	Crude oil, WTI	\\
UKX	&	Equities	&	FTSE 100 Index & CO1	&	Fuels	&	Crude oil, brent	\\
GDDUEMEA	&	Equities	&	EM: Europe, Middle East, Africa & HO1	&	Fuels	&	Heating oil	\\
GDUEEGFA	&	Equities	&	EM: Asia	& NG1	&	Fuels	&	Natural gas	\\
GDUEEGFL	&	Equities	&	EM: Latin America	& BO1	&	Commodities	&	Soybean oil	\\
ATGATR	&	Gov. bonds &	Austria & C 1	&	Commodities	&	Corn	\\
AUGATR	&	Gov. bonds &	Australia & CC1	&	Commodities	&	Cocoa	\\
BEGATR	&	Gov. bonds &	Belgium & CT1	&	Commodities	&	Cotton	\\
CAGATR	&	Gov. bonds &	Canada & FC1	&	Commodities	&	Coffee	\\
DEGATR	&	Gov. bonds &	Denmark & JN1	&	Commodities	&	Feeder cattle	\\
FIGATR	&	Gov. bonds &	Finland & JO1	&	Commodities	&	Orange juice	\\
FRGATR	&	Gov. bonds &	France & KC1	&	Commodities	&	Coffee	\\
GRGATR	&	Gov. bonds &	Germany & LB1	&	Commodities	&	Lumber	\\
IEGATR	&	Gov. bonds &	Ireland & LC1	&	Commodities	&	Live cattle	\\
ITGATR	&	Gov. bonds &	Italy & LH1	&	Commodities	&	Lean hogs	\\
JNGATR	&	Gov. bonds &	Japan & O 1	&	Commodities	&	Oats	\\
NEGATR	&	Gov. bonds &	Netherlands & PB1	&	Commodities	&	Frozen pork bellies	\\
NOGATR	&	Gov. bonds &	Norway & QW1	&	Commodities	&	Sugar	\\
NZGATR	&	Gov. bonds &	New Zealand & RR1	&	Commodities	&	Rough rice	\\
PTGATR	&	Gov. bonds &	Portugal & S 1	&	Commodities	&	Soybean	\\
SPGATR	&	Gov. bonds &	Spain & SM1	&	Commodities	&	Soybean meal	\\
SWGATR	&	Gov. bonds &	Sweden & W 1	&	Commodities	&	Wheat	\\
SZGATR	&	Gov. bonds &	Switzerland & MOODCAAA	&	Corp. bonds	&	Moody's AAA rated	\\
UKGATR	&	Gov. bonds &	U.K. & MOODCAA	&	Corp. bonds	&	Moody's AA rated	\\
USGATR	&	Gov. bonds &	U.S. & MOODCA	&	Corp. bonds	&	Moody's A rated	\\
AUDUSD	&	Currencies &	Australian dollar	& MOODCBAA	&	Corp. bonds	&	Moody's BAA rated	\\

\end{longtable*}
\end{center}

\section{Example Correlations Betweens Assets and Components}
\label{sec:egcorrs}

\begin{figure*}[htp]
\begin{center}
\includegraphics[width=1\linewidth]{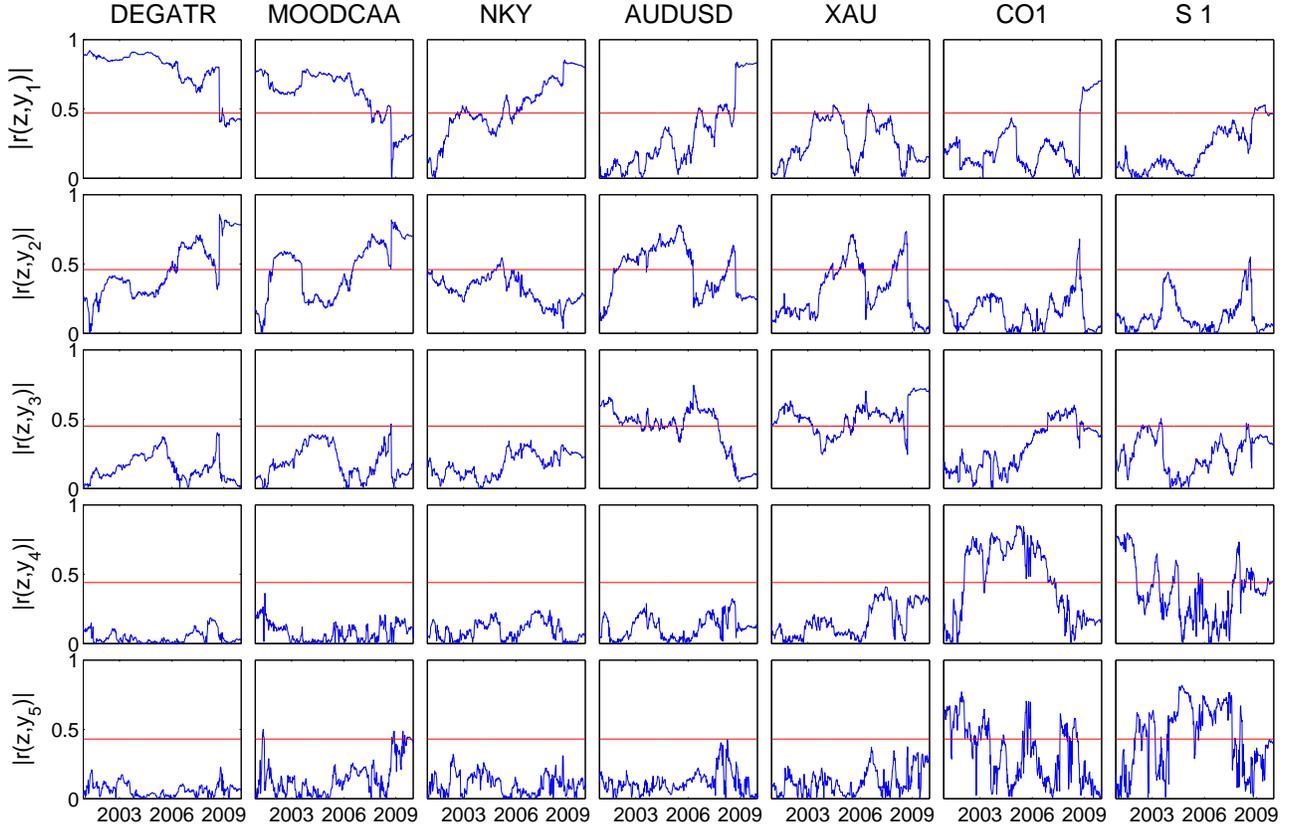}
\caption{Examples of the absolute correlation $\vert r(\hat{\mathbf{z}}_i,\mathbf{y}_k)\vert$ between various assets and the first five principal components ($k=1,\ldots,5$) as a function of time. We show Danish government bonds (DEGATR), AA-rated corporate bonds (MOODCAA), Nikkei 225 (NKY), gold (XAU), oil (CO1), and soybean futures (S 1). The first row shows correlations for the first PC, the second row for the second PC, and so on. The horizontal red lines show the values of the 99$^\textrm{th}$ percentile of the distribution of absolute correlation coefficients for the corresponding PC for random data. As we increase $k$ from 1 to 5, the 99$^\textrm{th}$ percentile of $\vert r(\hat{\mathbf{z}}_i,\mathbf{y}_k)\vert$ for random matrices decreases from 0.47 to 0.43.}
\label{fig:pc_corr_eg}
\end{center}
\end{figure*}

In Fig.~\ref{fig:pc_corr_eg}, we show example plots of the absolute correlation $\vert r(\hat{\mathbf{z}}_i,\mathbf{y}_k)\vert$ as a function of time for specific assets. These figures correspond to horizontal slices through the plots shown in Fig.~\ref{fig:sig_corr}. Figure~\ref{fig:pc_corr_eg} highlights that there are many time steps at which the absolute correlation between the PCs and asset return time series are significantly larger than the values for random matrices. For example, the correlation between Danish government bonds (DEGATR) and the first PC exceeds 0.9 at some time steps and is above the 99$^\textrm{th}$ percentile for random matrices at every time step until the collapse of Lehman Brothers. After Lehman's collapse, the correlation of DEGATR with the first PC falls sharply, but its correlation with the second PC increases to approximately 0.8. We observe similar behavior for AA-rated corporate bonds (MOODCAA), but there is also a period in 2003 during which corporate bonds were significantly correlated with the second PC. We can make similar observations for assets from other classes. For example, the absolute correlations of the AUDUSD exchange rate and oil futures (CO1) with the first PC are both significant after the collapse of Lehman Brothers; the correlation of gold (XAU) with the third PC is significant over the same period; and the correlation of the price of soybean futures (S 1) is initially significantly correlated with the fourth PC and then becomes significantly correlated with the first PC. Figure~\ref{fig:pc_corr_eg} demonstrates that by considering the changes in the magnitudes of the correlations between the PCs, we can gain insights into the changes taking place in markets.

\section{Contribution of Assets to the Correlations}
\label{sec:selfcorr}

The PCs are defined as linear combinations of the asset return time series $\hat{\mathbf{z}}_i$ [see Eq.~(\ref{eq:pcs})]. Because the assets contribute to the PCs, some of the correlation $\vert r(\hat{\mathbf{z}}_i,\mathbf{y}_k)\vert$ between the $i^\textrm{th}$ asset and the $k^\textrm{th}$ PC is attributable to asset $i$. To understand the effect of these self-correlations, we calculate the correlation between the return time series $\hat{\mathbf{z}}_i$ and each of the PCs with the contribution from the $i^\textrm{th}$ asset removed. For each asset $i$, we define $k=1,\ldots,N$ adjusted PCs $\mathbf{w}_k$ with elements given by
\begin{equation}
	w_k(t)=\sum_{i\neq k}^N\omega_{ki}\hat{z}_i(t)\
\end{equation}
and calculate the absolute correlation $\vert r(\hat{\mathbf{z}}_i,\mathbf{w}_k)\vert$ between each asset and the adjusted PCs.

\begin{figure}[htp]
\begin{center}
\includegraphics[width=1\linewidth]{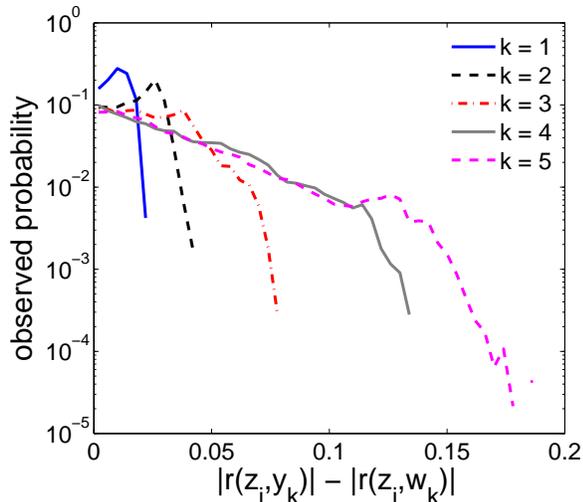}
\caption{(Color online) Distribution of the difference in absolute correlations $\vert r(\hat{\mathbf{z}}_i,\mathbf{y}_k)\vert$ between each asset and the first five PCs and the absolute correlations $\vert r(\hat{\mathbf{z}}_i,\mathbf{w}_k)\vert$ between each asset and the first five PCs with the contribution from the $i^\textrm{th}$ asset removed. For each component $k$, the distributions show the differences for all assets $i$.}
\label{fig:pc_var_corr_removed}
\end{center}
\end{figure}

In Fig.~\ref{fig:pc_var_corr_removed}, we compare the distribution of the differences between the absolute correlations $\vert r(\hat{\mathbf{z}}_i,\mathbf{y}_k)\vert$ (between the assets and the PCs) and the correlations $\vert r(\hat{\mathbf{z}}_i,\mathbf{w}_k)\vert$ (between the assets and the adjusted PCs). The correlations are very similar for $k=1$, which implies that the self-correlations do not dominate the correlations between PCs and assets. However, for larger $k$, the differences between $\vert r(\hat{\mathbf{z}}_i,\mathbf{y}_k)\vert$ and $\vert r(\hat{\mathbf{z}}_i,\mathbf{w}_k)\vert$ become more significant. This is expected because fewer assets make significant contributions to the higher PCs, so the removal of a single asset can have a larger effect. This highlights the fact that many assets contribute to the lower PCs, whereas the higher PCs are localized and have only a few assets contributing to them.

%%%%%%%%%%%%%%%%%%%%%%%%%%%%%%%%%%%%%%%%%%%%%%%%%%%%%%%%%%%%%%%%%%%

\end{document}